\documentclass[prd,twocolumn]{revtex4}

 \newcommand{\lan}{\langle}
 \newcommand{\ran}{\rangle}
 \newcommand{\be}{\begin{equation}}
 \newcommand{\bea}{\begin{eqnarray}}
 \newcommand{\eea}{\end{eqnarray}}
 \newcommand{\ee}{\end{equation}}

\def\la{\mathrel{\mathpalette\fun <}}

\def\fun#1#2{\lower3.6pt\vbox{\baselineskip0pt\lineskip.9pt
\ialign{$\mathsurround=0pt#1\hfil##\hfil$\crcr#2\crcr\sim\crcr}}}

\begin{document}
\title{Low temperature relation for the trace\\ of the
energy-momentum tensor in QCD with light quarks}
\author{N.O.Agasian}
\email{agasian@heron.itep.ru}
\affiliation{State Research Center\\
Institute of Theoretical and Experimental Physics,\\
Moscow 117218, Russia}

\begin{abstract}
It is shown that the temperature derivatives of the anomalous and normal
(quark massive term) contributions to the trace of the energy-momentum tensor
in  QCD  are equal to each other in the low temperature region.
The physical consequences of this relation are discussed.
\end{abstract}
\pacs{11.10.Wx,12.38.Aw,12.38.Mh}

\maketitle
\newpage

 1. The low-energy theorems, playing an important role in the understanding
 of the vacuum state properties in quantum field theory,
were discovered almost at the same time as quantum field methods have been applied
in particle physics (see, for example, Low theorems \cite{low}).
In QCD, they were obtained in the beginning of eighties
\cite{LT}. The QCD low-energy theorems, being derived from the very general
symmetry considerations and not depending on the details
of confinement mechanism, sometimes give information which
is not easy to obtain in another way. Also, they can be used
as "physically sensible" restrictions in the constructing
of effective theories. Recently, they were generalized to
finite temperature and chemical potential case \cite{k1,k2}.
These theorems were used for investigation of QCD vacuum phase
structure in a magnetic field \cite{a1} and at finite
temperature \cite{a2}.

 The investigation of the vacuum state behavior under the
 influence of various external factors is known to be one of the
 central problems of quantum field theory. In the realm of strong
 interactions (QCD) the main factors are the temperature and
 the baryon density. At low temperatures, $T<T_c$ ( $T_c$--temperature of the
 "hadron--quark-gluon" phase transition ), the dynamics of QCD
 is essentially nonperturbative and
 is characterized by confinement and spontaneous
breaking of chiral symmetry (SBCS). In the hadronic phase the
partition function of the system is dominated by the contribution
 of the lightest  particles in the physical spectrum.
It is well known that due to the smallness of pion mass as
compared to the typical scale of strong interactions, the pion
plays a special role among other strong-interacting particles.
Therefore for many problems of QCD at zero temperature the chiral
limit, $M_\pi\to 0$, is an appropriate one. On the other hand a
new mass scale emerges in the physics of QCD phase transitions,
namely the critical transition temperature $T_c$.
Numerically the critical transition temperature turns out to be
close to the pion mass, $T_c\approx M_\pi$ \cite{kar}.
However hadron states heavier than pion have masses several
times larger than $T_c$ and therefore their contribution
to the thermodynamic quantities is damped
by Boltzmann factor $\sim \exp \{-M_{hadr}/T\}$.
Thus the thermodynamics of the low temperature
hadron phase, $T\la M_\pi$, is described basically in terms of the
thermal  excitations of relativistic massive pions.

In the present paper the low temperature relation
for the trace of the energy-momentum tensor in QCD
with two light quarks is obtained based on the general dimensional and
renormalization-group properties of the QCD partition function
and dominating role of the pion thermal exitations in the hadronic phase.
The physical consequences of this relation are discussed
as well as the possibilities to use it in the lattice studies
of the QCD at finite temperature.

 2. For non-zero quark mass ($m_q\neq 0)$ the scale invariance is
 broken already at the classical level.
Therefore the pion thermal  excitations would change, even in the
ideal gas approximation, the value of the gluon condensate
with increasing temperature \cite{ft1}.
 To determine this dependence use will be made of the general
 renormalization and scale properties of the QCD partition
 function. This is a standard method  and it is used for derivation of
 low-energy QCD theorems [2-6]. In Ref. \cite{agas} this method was
 used for investigation of the thermodynamic properties
 of QCD nonperturbative vacuum  with two flavors  at low
 temperature outside of the scope  of perturbation theory.
 In what follows we stick to the approach outlined  in \cite{agas}.

  The QCD Euclidean partition function with two quark flavors
 has the following form ($\beta=1/T$)
 \be Z= \int[DA] \prod_{q=u,d} [D\bar q][Dq]
\exp \left \{ -\int^\beta_0 dx_4\int_V d^3x {\cal L} \right \}.
\label{2}
 \ee
  Here the QCD Lagrangian  is
  \be
 {\cal L}=\frac{1}{4g^2_0}
 (G^a_{\mu\nu})^2+ \sum_{q=u,d} \bar q[\gamma_\mu
 (\partial_\mu-i\frac{\lambda^a}{2} A^a_\mu)+m_{0q}]q,
 \label{3}
  \ee
where the gauge fixing and ghost terms have been omitted. The
free energy density is given by the relation $ \beta VF$
$(T,m_{0u},m_{0d})=-\ln Z$.  Eq. (\ref{2}) yields
the following expression for the gluon condensate
($\langle G^2\rangle\equiv \langle (G^a_{\mu\nu})^2\rangle$)
  \be
    \langle G^2\rangle (T,m_{0u},m_{0d})=4\frac{\partial F}{\partial(1/g^2_0)}.
 \label{4}
 \ee
 The system described by the partition function (\ref{2}) is
 characterized by the set of dimensionful parameters $M, T, m_{0q}
 (M)$ and dimensionless charge $g^2_0(M)$, where $M$ is the
 ultraviolet cutoff.
 On the other hand one can consider the renormalized free energy
 $F_R$ and by using the dimensional and renormalization-group properties
 of $F_R$ recast (\ref{4}) into the form containing derivatives
 with respect to the physical parameter $T$ and renormalized masses $m_q$.

The phenomenon of dimensional transmutation results in the
appearance of a nonperturbative dimensionful parameter
 \be
  \Lambda= M \exp
 \left \{ \int^\infty_{\alpha_s(M)}\frac{d\alpha_s}{\beta(\alpha_s)}
 \right \}~,
  \label{5}
  \ee
  where
  $\alpha_s=g^2_0/4\pi$, and $\beta(\alpha_s)=d\alpha_s(M)/d
  ~ln M$  is the Gell-Mann-Low  function.
Furthermore, as it is well known,    the quark mass has anomalous
dimension and depends on the scale $M$. The  renormalization
-group equation for $m_0(M)$, the running mass, is  $d\ln m_0/d\ln
M=-\gamma_m$ and we use the $\overline{MS} $ scheme for which
$\beta$ and $\gamma_m$ are independent of the quark mass
\cite{k2,muta}. Upon integration the renormalization-group invariant
mass is given by
\be
m_q=m_{oq}(M)\exp\{\int^{\alpha_s(M)}\frac{\gamma_{m_q}(\alpha_s)}{\beta(\alpha_s)}
d\alpha_s\},
 \label{6}
 \ee
  where the indefinite integral is
evaluated at $\alpha_s(M)$. Next we note that since free energy is
renormalization-group invariant quantity its anomalous dimension
is zero. Thus $F_R$ has only a normal (canonical) dimension equal
to 4. Making use of the renorm-invariance of $\Lambda$, one can
write in the most general form
\be
F_R=\Lambda^4 f(\frac{T}{\Lambda}, \frac{m_u}{\Lambda},
\frac{m_d}{\Lambda}),
 \label{7}
 \ee
 where $f$ is some function.
 From (\ref{5}),(\ref{6}) and (\ref{7}) one gets
\be
\frac{\partial F_R}{\partial(1/g^2_0)}=
 \frac{\partial
F_R}{\partial\Lambda} \frac{\partial\Lambda}{\partial(1/g^2_0)} +
\sum_q \frac{\partial F_R}{\partial m_q} \frac{\partial
m_q}{\partial(1/g^2_0)},
 \label{8}
 \ee

\be \frac{\partial m_q}{\partial(1/g^2_0)}=-4\pi\alpha^2_s
m_q\frac{\gamma_{m_q}(\alpha_s)}{\beta(\alpha_s)}.
 \label{9}
 \ee
 With the account of (\ref{4}) the gluon condensate is given by
$$
\lan G^2\ran (T, m_u, m_d)
$$
\be
=\frac{16\pi\alpha_s^2}{\beta(\alpha_s)}(4-T\frac{\partial}{\partial
T}-\sum_q(1+\gamma_{m_q})m_q\frac{\partial}{\partial {m_q}}) F_R.
\label{10}
 \ee
 It is convenient to choose such a large scale that one can take the
lowest order expressions,  $\beta(\alpha_s)\to -
b\alpha^2_s/2\pi$, where $b=(11 N_c-2N_f)/3$ and $1+\gamma_m\to
1$. Thus, we have the following equations for condensates
\be
\lan G^2\ran (T)=-\frac{32\pi^2}{b} (4-T\frac{\partial}{\partial
T}-\sum_q m_q\frac{\partial}{\partial m_q}) F_R\equiv
 -\hat DF_R~,
 \label{11}
 \ee
 \be
 \lan\bar q q\ran (T)=\frac{\partial F_R}{\partial {m_q}}~.
 \label{12}
 \ee

 3. In the hadronic phase the effective pressure from which one can extract the
condensates $\lan \bar q q\ran(T)$ and $\lan G^2\ran(T)$ using
the general relations (\ref{11}) and (\ref{12}) has the form
\be
P_{eff}(T)=-\varepsilon_{vac}+P_h(T),
\label{14}
\ee
where
$\varepsilon_{vac}=\frac14\lan\theta_{\mu\mu}\ran$ is the
nonperturbative vacuum energy density at $T=0$ and
\be
\lan
\theta_{\mu\mu}\ran=-\frac{b}{32\pi^2} \lan G^2\ran+\sum_{q=u,d}
m_q\lan\bar qq \ran
\label{15}
\ee
is the trace of the
energy-momentum tensor. In Eq.(\ref{14}) $P_h(T)$ is the
thermal hadrons pressure.
The quark and gluon condensates are given by the equations
\be
\lan \bar qq\ran
(T)=-\frac{\partial P_{eff}}{\partial m_q},
\label{17}
\ee
\be
\lan G^2\ran (T)= \hat DP_{eff},
\label{18}
\ee
where the
operator $\hat D$ is defined by the relation (\ref{11})
\be \hat
D=\frac{32\pi^2}{b} (4-T\frac{\partial}{\partial T}-\sum_q
m_q\frac{\partial}{\partial m_q}).
\label{19}
\ee

Consider the $T=0$ case. One can use the low energy theorem
for the derivative of the gluon condensate with respect to the
quark mass \cite{LT}
\be
\frac{\partial}{\partial m_q}\lan G^2\ran= \int d^4 x\lan G^2(0)
\bar q q(x)\ran =-\frac{96\pi^2}{b}\lan \bar q q\ran+O(m_q),
\label{20}
\ee
where $O(m_q)$ stands for the terms linear in light
quark masses.Then one arrives at the following
relation \cite{ft2}
\be
\frac{\partial\varepsilon_{vac}}{\partial m_q}=-
\frac{b}{128\pi^2}\frac{\partial}{\partial m_q} \lan
G^2\ran+\frac{1}{4}\lan \bar q q\ran =\frac34 \lan \bar q
q\ran+\frac14\lan \bar q q\ran=\lan \bar q q\ran.
\label{21}
\ee
Note that three fourths of the quark condensate stem from the
gluon part of the nonperturbative vacuum energy density. Along the
same lines one arrives at the expression for the gluon
condensate
\be
-\hat D\varepsilon_{vac}=\lan G^2\ran.
\label{22}
\ee

In order to get the dependence of the quark and
gluon condensates upon $T$ use is made of the Gell-Mann-
Oakes-Renner (GMOR) relation ($\Sigma=|\lan\bar u u\ran|=|\lan
\bar dd\ran|$)
\be F^2_\pi M^2_\pi=-\frac12(m_u+m_d)\lan \bar
uu+\bar dd\ran=(m_u+m_d)\Sigma
\label{23}
\ee
Then we can find the following relations
\be
\frac{\partial}{\partial
m_q}=\frac{\Sigma}{F^2_\pi} \frac{\partial}{\partial M^2_\pi},
\label{24}
\ee
\be \sum_qm_q\frac{\partial}{\partial
m_q}=(m_u+m_d)\frac{\Sigma}{F^2_\pi}\frac{\partial}{\partial
M^2_\pi}=M^2_\pi\frac{\partial}{\partial M^2_\pi},
\label{25}
\ee
\be \hat D=\frac{32\pi^2}{b}(4-T\frac{\partial}{\partial
T}-M^2_\pi\frac{\partial}{\partial M^2_\pi}).
\label{26}
\ee

 Within the described above framework one can derive the thermodynamic
relation for the quantum anomaly in the trace
 of the energy-momentum tensor of QCD. At low temperature the main
 contribution to the pressure comes from thermal excitations of
 massive pions. The general expression for the pressure reads
 \be
 P_\pi=T^4\varphi(M_\pi/T),
 \label{36}
 \ee
 where $\varphi$ is a function of the ratio $M_\pi/T$.
 Then the following  relation is valid
 \be
 (4-T\frac{\partial}{\partial T}-M^2_\pi\frac{\partial}{\partial
 M^2_\pi}) P_\pi=M^2_\pi\frac{\partial P_\pi}{\partial M^2_\pi}.
 \label{37}
 \ee
With the account of (\ref{17},\ref{18}), (\ref{21},{22}) and (\ref{37})
one gets
\be
\Delta \lan \bar qq\ran =-\frac{\partial P_\pi}{\partial m_q},~~
\Delta \lan G^2\ran=\frac{32\pi^2}{b} M^2_\pi\frac{\partial
P_\pi}{\partial M^2_\pi},
\label{38}
\ee
where $ \Delta \lan \bar
qq\ran= \lan \bar qq\ran_T- \lan \bar qq\ran $ and $\Delta \lan
G^2\ran=  \lan G^2\ran_T- \lan G^2\ran.$ In view of (\ref{25}) one
can recast (\ref{38}) in the form
\be
\Delta \lan G^2\ran=-\frac{32\pi^2}{b} \sum_q m_q\Delta  \lan \bar
qq\ran
\label{39}
\ee
Let us divide both sides of (\ref{39}) by
$\Delta T$ and take the limit $\Delta T\to 0$. This yields
\be
\frac{\partial \lan G^2\ran}{\partial T}=-\frac{32\pi^2}{b} \sum_q
m_q\frac{\partial\lan \bar qq\ran}{\partial T}.
\label{40}
\ee
This can be rewritten as
\be
\frac{\partial \lan \theta^g_{\mu\mu}\ran}{\partial
T}=\frac{\partial \lan \theta^q_{\mu\mu}\ran}{\partial T}
\label{41}
\ee
where $\lan \theta^q_{\mu\mu}\ran=\sum m_q \lan
\bar qq\ran$ and $\lan
\theta^g_{\mu\mu}\ran=(\beta(\alpha_s)/16\pi\alpha^2_s) \lan
G^2\ran$ are correspondingly the quark and gluon contributions
to the trace of the energy-momentum tensor.
Note that in deriving this result use was made of
the low energy GMOR relation, and therefore the thermodynamic
relation (\ref{40},\ref{41}) is valid in the light quark theory.

In Ref.\cite{agas} it was shown that dilute gas approximation for the
relativistic massive pions is valid at  temperature $T\la M_\pi$.
The pressure of the massive relativistic pion gas has the form
\be
P_\pi(T)
 =\frac{3M^2_\pi T^2}{2\pi^2}
\sum^\infty_{n=1}\frac{1}{n^2}K_2(n\frac{M_\pi}{T}),
\label{27}
\ee
where $K_2$ is the Mackdonald function.
Making use of the relations (14-23) and pressure (\ref{27})
we find the following expressions for the condensates
$\Sigma(T)$ and $\lan G^2\ran (T)$ \cite{agas}
\be
\frac{\Sigma(T)}{\Sigma}=1-\frac{3M_\pi
T}{4\pi^2F^2_\pi}\sum^\infty_{n=1}
\frac{1}{n}K_1(n\frac{M_\pi}{T}).
\label{28}
\ee

\be
\frac{\lan G^2\ran
(T)}{\lan G^2\ran }=1
-\frac{24}{b}\frac{M^3_\pi T}{\lan G^2\ran
}\sum^\infty_{n=1}\frac{1}{n}K_1(n\frac{M_\pi}{T}).
\label{31}
\ee
Analytic temperature dependence of the quark condensate (\ref{40})
perfectly agrees, in the low temperature region, $T\la M_\pi$,
with the numerical  calculations of $\lan \bar q q\ran(T)$
obtained at the three-loop level of the chiral perturbation theory (ChPT)
with non-zero quark mass \cite{GL}.
The gluon condensate slightly varies with the increase of the temperature
($\Delta\lan G^2\ran/\lan G^2\ran \sim 10^{-3}$
at $T=M_\pi$  and $\lan G^2\ran =0.5$ GeV$^4$),
i.e. the situation is similar to the ChPT \cite{Leut}.

Turning back, the relation (\ref{40},\ref{41}) can be easily verified via direct
calculation using (\ref{28}), (\ref{31}) and the GMOR relation.
Thus in the low temperature region when the excitations of massive
hadrons and interactions of pions can be neglected, equation
(\ref{41}) becomes a rigorous QCD theorem.

As it was mentioned above the pion plays an exceptional role in
thermodynamics of QCD due to the fact that its mass is numerically
close to the phase transition temperature while the masses of
heavier hadrons are several times larger than $T_c$. This was the
reason we did not consider the role of massive states  in the low
temperature phase. This question was discussed in detail in
Ref.\cite{GL}. It was shown there that at low temperatures, the
contribution to $\lan \bar q q\ran$ generated by the massive states
is very small, less than 5\% if $T$ is below 100 MeV. At $T=150$ MeV,
this contribution is of the order of 10\%. The influence of
thermal excitations of massive hadrons on the properties of the
gluon and quark condensates in the framework of the
conformal-nonlinear $\sigma$- model was also studied in detail in
\cite{AEI}.

4. It was shown that the temperature derivatives of the
anomalous and normal (quark massive term) contributions to the
trace of the energy-momentum tensor in  QCD with light quarks are
equal to each other in the low temperature region.

Let us consider some physical consequences and  possible
applications of this relation. To this end we introduce the
function

\be
\delta_\theta(T)={\frac \partial {\partial T} \langle \theta^g_{\mu\mu}}
-\theta^q_{\mu\mu}\rangle
\label{dt}
\ee
 As it was stated above, the function $\delta_\theta(T)$ at low
temperatures is, with  good accuracy, close to zero. In the
vicinity and at the phase transition point, i.e. in the region of
nonperturbative vacuum  reconstruction this function changes
drastically. To see it, we first consider pure gluodynamics. It
was shown in \cite{dilat} using the effective dilaton Lagrangian,
that gluon condensate decreases very weakly with the increase of
temperature, up to phase transition point. This result is
physically transparent and is the consequence of Boltzmann
suppression of thermal glueball excitations in the confining
phase.

Further, in Refs.\cite{sim} the dynamical picture of deconfinement
was suggested based on the reconstruction of the nonperturbative
gluonic vacuum. Namely, confining and deconfining phases according to
\cite{sim} differ first of all in the vacuum fields, i.e., in the value of
the gluon condensate and in the gluonic field correlators. It was
argued in \cite{sim} that color-magnetic (CM) correlators and their
contribution to the condensate are kept intact across the temperature phase
transition, while the confining color-electric (CE) part abruptly disappears
above $T_c$. Furthermore, there exist numerical lattice measurements
of field correlators near the critical transition temperature $T_c$,
made by the Pisa group \cite{dig1}, where both CE and CM correlators
are found with good accuracy. These data clearly demonstrate
the strong suppression of CE  component above $T_c$ and persistence
of CM components.
Thus, the function $\delta_\theta(T)_{GD} =\partial\langle \theta^g_{\mu\mu}\rangle/\partial
T$ can be presented as a
$\delta$-function smeared around the critical point $T_c$ with the
width $\sim \Delta T$ which defines the fluctuation region of phase
transition.

Similar, but more complicated and interesting situation takes
place in the theory with quarks. The function $\delta_\theta(T)$
contains the quark term, proportional to the chiral phase
transition order parameter $\langle \bar qq\rangle (T)$. So it is
interesting to check the relation (\ref{41}) and to study the
behavior of the function $\delta_\theta(T)$  in the lattice QCD
at finite  temperature. It would allow both to test the
nonperturbative QCD vacuum at the low temperatures in the
confining phase and to extract additional information on the
thermal phase transitions in QCD.
\begin{center}

{\bf ACKNOWLEDGMENTS} \\
\end{center}

I am grateful to Yu.A.Simonov for useful discussions and comments.
The financial support of RFFI grant 00-02-17836 and
INTAS grant CALL 2000 N 110 is gratefully acknowledged.


\begin{thebibliography}{99}
\bibitem{ft1}
At zero quark mass
 the gas of massless noninteracting pions is obviously
 scale-invariant and therefore does not contribute to
the trace of the  energy-momentum tensor and correspondingly to the
gluon condensate $\langle (G^a_{\mu\nu})^2\rangle$.
\bibitem{ft2}
The equation (\ref{21}) contains corrections
$\sim  m_q\partial\lan \bar qq\ran/\partial m_q \sim O(m_q)$
which are negligible for $u$ and $d$ quarks.
\end{thebibliography}
\end{document}